# MAGNETIC LATTICES FOR ULTRACOLD ATOMS *

TIEN DUY TRAN[a], YIBO WANG[b], ALEX GLAETZLE[c], SHANNON WHITLOCK[b], ANDREI SIDOROV[a] AND PETER HANNAFORD[a]

[a]*Centre for Quantum and Optical Science, Swinburne University of Technology, Melbourne, Victoria 3122, Australia*

[b]*IPCMS (UMR 7504) and ISIS (UMR 7006), University of Strasbourg and CNRS, 67000 Strasbourg, France*

[c]*Clarendon Laboratory, University of Oxford, Parks Road, Oxford OX1 3PU, United Kingdom*

*E-mail*: phannaford@swin.edu.au

*8 March 2019*

**Abstract.** *This article reviews the development in our laboratory of magnetic lattices comprising periodic arrays of magnetic microtraps created by patterned magnetic films to trap periodic arrays of ultracold atoms. Recent achievements include the realisation of multiple Bose-Einstein condensates in a 10 µm-period one-dimensional magnetic lattice; the fabrication of sub-micron-period square and triangular magnetic lattice structures suitable for quantum tunnelling experiments; the trapping of ultracold atoms in a sub-micron-period triangular magnetic lattice; and a proposal to use long-range interacting Rydberg atoms to achieve spin-spin interactions between sites in a large-spacing magnetic lattice.*

## I. INTRODUCTION

Following the discovery in the mid-1980s of laser techniques to cool and trap clouds of atoms down to microkelvin temperatures [1-3], optical lattices – periodic arrays of optical dipole traps created by interfering laser beams – have become a standard tool for trapping arrays of ultracold atoms and degenerate quantum gases [4-6]. Applications of optical lattices include quantum simulations of condensed matter phenomena [5-7], super-precise optical lattice clocks [8], and quantum gates for quantum information processing [9, 10]. Such lattices allow precise control over lattice parameters, such as the lattice geometry, the lattice spacing, and the inter-particle interaction, and provide an ideal platform to realise a variety of exotic condensed matter phenomena (e.g., [5, 6]). Examples include the realisation of the superfluid to Mott insulator transition in the Bose-Hubbard model [11], antiferromagnetic correlations between fermionic atoms in the Hubbard model [12], antiferromagnetic spin chains of atoms in the Ising model [13], low-dimensional bosonic [14] and fermionic [15] systems, topological insulators involving edge states [16], and Josephson junction arrays [17].

An alternative approach for creating periodic lattices of ultracold atoms, which we have been investigating in Melbourne, involves the use of periodic arrays of magnetic microtraps created by patterned magnetic films on an atom chip [18-41]. Such magnetic lattices offer a high degree of design flexibility and may, in principle, be tailored with nearly arbitrary configurations and lattice spacing [27], and they may be readily scaled up to millions of lattice sites. In addition, magnetic lattices do not require (high-power) laser beams and precise beam alignment, they operate with relatively little technical noise and heating, and they provide state-selective atom trapping, allowing radiofrequency (RF) evaporative cooling to be performed in the lattice and RF spectroscopy to characterise the trapped atoms *in situ* [42, 43]. Finally, magnetic lattices have the potential to enable miniaturized integrated quantum technologies such as 'atomtronics' [44].

We have recently shown that magnetic lattices can be used to realise multiple Bose-Einstein condensates in a 10 $\mu$m-period magnetic lattice [33, 35] near the transition to the quasi-one-dimensional regime, thus opening the possibility to study ultracold quantum matter in different geometries. For many applications, however, it would also be desirable to have interactions between the atoms on neighbouring sites, for example, via quantum tunnelling. For magnetic lattices, achieving significant tunnelling between neighbouring sites presents a challenge, since the lattice spacing needs to be in the sub-micron range which also requires the fabrication of sub-micron magnetic structures and trapping of the atoms at sub-micron distances from the chip surface. We have recently shown that suitable sub-micron-period square and triangular magnetic lattice structures for quantum tunnelling experiments can be fabricated by patterning Co/Pd multi-layered magnetic films [36]. We also demonstrated the trapping of ultracold atoms in a 0.7 µm-period triangular magnetic lattice at distances down to about 100 nm from the chip surface [38]. At these distances, however, losses due to surface effects can be problematic. Possible surface effects at sub-micron distances include the Casimir-Polder interaction [45], Stark shifts due to electric fields created by atoms adsorbed onto the chip surface during each cooling and trapping sequence [46], and transitions between Zeeman sublevels (spin flips) due to magnetic Johnson noise [45, 47, 48]. An alternative approach is to increase the atom-atom interactions between distant lattice sites, for example, by exciting the atoms on neighbouring sites to long-range interacting Rydberg states. The advantage of Rydberg states is that their van der Waals inter-particle interaction energy scales as $n^{11}$ with the principal quantum number $n$, enabling strong interactions that can extend over several micrometers [49]. For this reason we recently put forward a proposal to use long-range interacting Rydberg atoms to realise spin-spin interactions between sites in a large-period magnetic lattice for simulating frustrated quantum spin models [50]. However, increasing the range of the atom-atom interaction also increases the range of the atom-surface interaction, and this can also present new challenges [51, 52].

In the following, we summarise our basic approach to trapping atoms using magnetic lattices, the design and fabrication of sub-micron-period magnetic lattices, investigations of atom trapping in sub-micron-period magnetic lattices and studies of the trapping lifetime due to surface effects, and finally we review our proposal for simulating quantum spin models using Rydberg states prepared in magnetic lattices.

## II. TRAPPING ULTRACOLD ATOMS IN A 10-µm-PERIOD ONE-DIMENSIONAL MAGNETIC LATTICE

A one-dimensional periodic array of magnets provides the simplest example of the magnetic lattice concept and is a starting point for more complex (two-dimensional) lattice geometries.

For an infinite one-dimensional periodic array of long magnets (in the *x-y* plane) with perpendicular magnetisation $M_z$, period *a* and bias fields $B_x$, $B_y$, the magnetic field components at distances $z \gg a/2\pi$ from the bottom of the magnetic surface are given approximately by [18]

$$[B^x; B^y; B^z] \approx [B_x;\ B_0 \sin(ky)\, e^{-kz} + B_y;\ B_0 \cos(ky)\, e^{-kz}], \quad (1)$$

where $k = 2\pi/a$, $B_0 = 4M_z(e^{kt} - 1)$ is a characteristic surface magnetic field (in Gaussian units), and $t$ is the thickness of the magnets. The magnetic field minimum $B_{\min}$ (or trap bottom), trapping height $z_{\min}$, barrier heights $\Delta B_{y,z}$, and trap frequencies $\omega_{y,z}$ for atoms of mass $m$ in a harmonic trapping potential are given by [18]

$$B_{\min} = |B_x| \quad (2)$$

$$z_{\min} = \frac{a}{2\pi} \ln\left(\frac{B_0}{|B_y|}\right) \quad (3)$$

$$\Delta B_y = (B_x^2 + 4B_y^2)^{1/2} - |B_x|; \quad \Delta B_z = (B_x^2 + B_y^2)^{1/2} - |B_x| \quad (4)$$

$$\omega_y = \omega_z = \omega_{\rm rad} = \frac{2\pi}{a}\left(\frac{m_F g_F \mu_B}{m|B_x|}\right)^{1/2} |B_y|, \quad (5)$$

where $m_F$ is the magnetic quantum number, $g_F$ is the Landé g-factor and $\mu_B$ is the Bohr magneton. Thus $B_{\min}$, $z_{\min}$, $\Delta B_{y,z}$ and $\omega_{y,z}$ may be controlled by adjusting the bias fields $B_x$ and $B_y$.

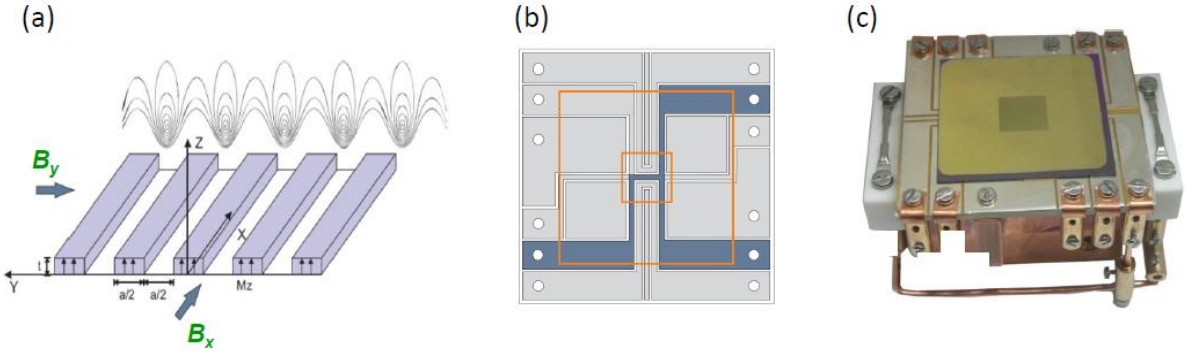

**Fig. 1.** (a) Magnetic lattice of a 1D array of magnetic microtraps created by a periodic array of perpendicularly magnetised magnets with period $a$ and bias fields $B_x$, $B_y$. The contour lines are equipotentials calculated for typical parameters. (b) Conducting film beneath the magnetic lattice structure comprising U-shaped and Z-shaped current-carrying conductors for trapping and loading the atoms. (c) Photograph of the assembled magnetic lattice chip coated with a reflecting gold film. Figures adapted from [21, 53].

Our first experiments were performed on a 10 µm-period one-dimensional magnetic lattice created by a perpendicularly magnetised TbGdFeCo film deposited on a grooved silicon substrate on an atom chip plus bias fields (Fig. 1) [21, 53]. Rubidium-87 atoms were initially trapped in a mirror magneto-optical trap (MOT) and then confined in a compressed MOT using the quadrupole field from a current-carrying U-wire on the atom chip plus bias field. The atoms were then pumped into the $|F = 2, m_F = +2\rangle$ low-field-seeking state and transferred to a Z-wire magnetic trap (with non-zero $B_{\min}$) and evaporatively cooled down to ~15 µK. The Z-wire trap was then brought close (~5 µm) to the surface of the atom chip by ramping down the Z-wire current ($I_z$) and ramping up the bias field $B_y$ to create a 1D lattice of magnetic microtraps. When the Z-wire trap merged with the magnetic lattice traps, $I_z$ was reduced to zero with $B_{min}$=15 G. In this way ~$10^6$ atoms were loaded into ~100 elongated magnetic lattice traps with barrier heights of ~1 mK and trap frequencies of $\omega_{rad}/2\pi$=20-90 kHz, $\omega_{ax}/2\pi \approx 1$ Hz [21]. *In situ* RF spectroscopy measurements indicated atom temperatures of >150 µK, which were limited by the weak axial confinement.

In the next generation of experiments [33, 35], stronger axial confinement was employed, with lattice trap frequencies of $\omega_{rad}/2\pi = 1.5 - 20$ kHz, $\omega_{ax}/2\pi = 260$ Hz. This allowed the atoms to be evaporatively cooled to much lower trap depths $\delta f = f_f - f_0$ (where $f_f$ and $f_0$ are the final evaporation frequency and trap bottom) since atoms satisfying the resonance condition $hf = \mu B$ are outcoupled from the traps, and hence to be cooled to lower temperatures [33]. In addition, the Rb atoms were prepared in the $|F = 1, m_F = -1\rangle$ state which has a smaller three-body recombination rate than the $|F = 2, m_F = 2\rangle$ state [54, 55]. Site-resolved RF spectra taken for about 100 lattice sites in the central region of the lattice revealed the evolution from an initial broad thermal cloud distribution (Fig. 2(a)) to a bimodal distribution (Fig. 2(b)) to an almost pure Bose-Einstein condensate distribution (Fig. 2(c)) as the atom clouds were cooled through the critical temperature (1.6 µK for an ideal gas with $N$ = 3000 atoms/site).

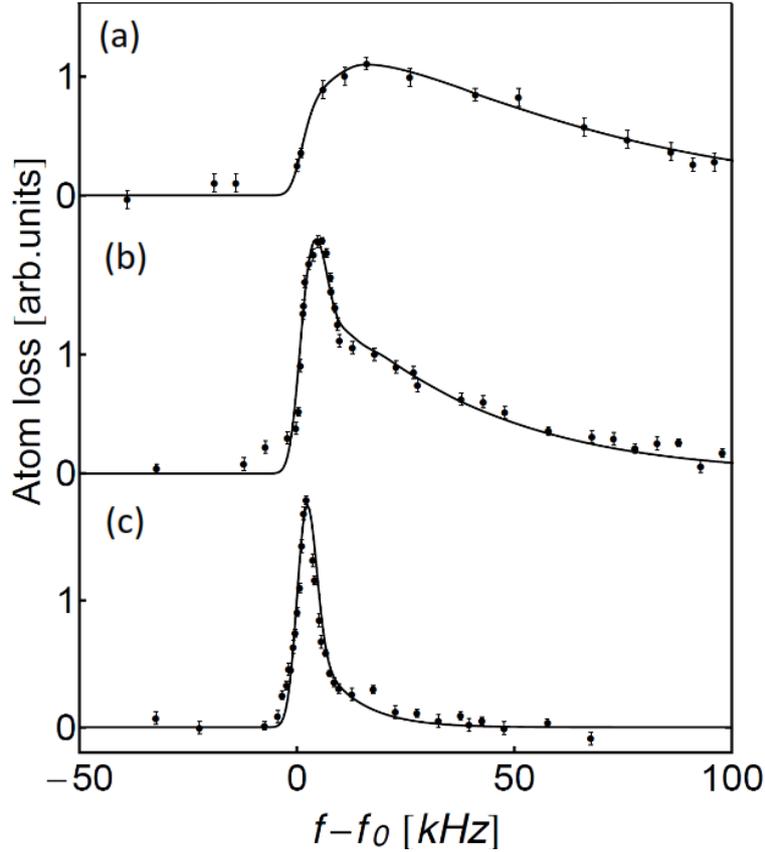

**Fig. 2.** Radiofrequency spectra of $^{87}$Rb $|F = 1, m_F = -1\rangle$ atoms in one of the ~100 atom clouds trapped in a 10 μm-period 1D magnetic lattice, demonstrating the onset of Bose-Einstein condensation with increased evaporative cooling. The solid lines are fits to the data based on a self-consistent mean-field model for a BEC plus thermal cloud [35]. The temperatures and atom numbers obtained from this analysis are (a) 2.0 μK, 5350 atoms (b) 1.3 μK, 3430 atoms and (c) 0.38 μK, 200 atoms. Adapted from [35].

Radiofrequency spectra taken simultaneously for ~100 atom clouds across the central region of the magnetic lattice showed similar bimodal distributions to Fig. 2 with site-to-site variations in the trap bottom $f_0$, , the atom temperature $T$, the condensate fraction $N_C/N$ and the chemical potential μ that were within the measurement uncertainties (Fig. 3). In particular, the trap bottom, which could be precisely determined from the frequency at which there were no atoms remaining (Fig. 2(c)), showed variations of only ± 0.3 kHz in 5 MHz (Fig. 3(b)), reflecting the high degree of uniformity in the central region of the magnetic lattice.

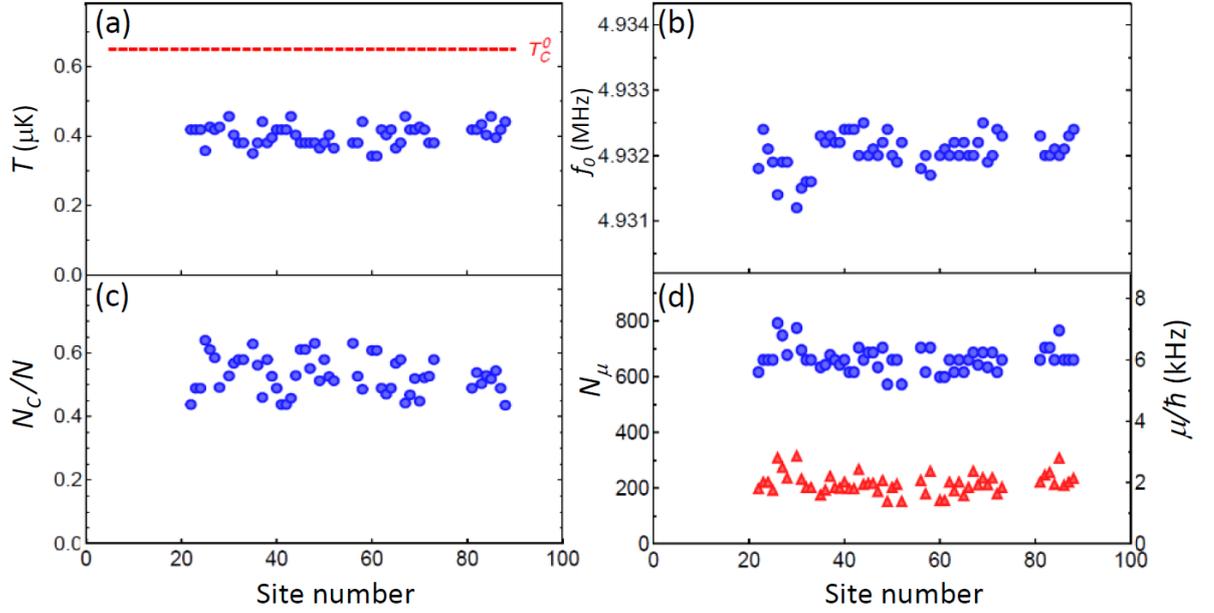

**Fig. 3.** (a) Atom temperature $T$, (b) trap bottom $f_0$, (c) condensate fraction $N_C/N$ and (d) chemical potential $\mu/\hbar$ (blue circles) and atom number $N_\mu$ (red triangles), determined from fits to the RF spectra for 54 sites across the central region of the magnetic lattice. The red dashed line in (a) represents the ideal-gas critical temperature for 220 atoms. Adapted from [35].

At the smallest trap depth (50 kHz), a temperature of 0.25 μK is achieved in the magnetic lattice (Fig. 4(a)) with a condensate fraction of 81% (Fig. 4(b)), while at the lowest radial trap frequency (1.5 kHz) a temperature of 0.16 μK is achieved (Fig. 4(c)). For $\omega_{rad}/2\pi > 10$ kHz, both the chemical potential μ and the thermal energy $k_B T$ become smaller than the energy of the lowest radial vibrational excited state $\hbar\omega_{\text{rad}}$ (Fig. 4(d)), which represents the quasi-one-dimensional Bose gas regime [14, 56, 57].

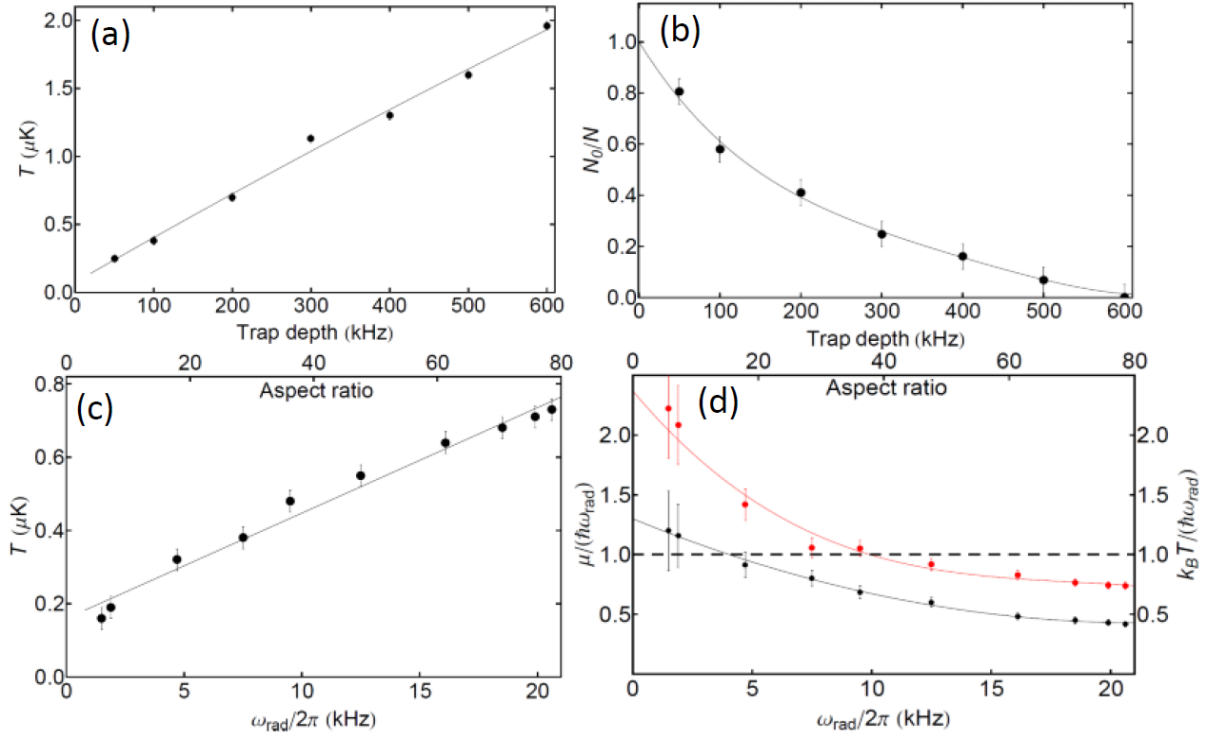

**Fig. 4.** Variation of (a) atom temperature $T$ and (b) condensate fraction $N_C/N$ with trap depth $\delta f = f_f - f_0$ at trap frequencies $\omega_{rad}/2\pi$ = 7.5 kHz, $\omega_{ax}/2\pi$ = 260 Hz; and variation of (c) atom temperature and (d) the ratios $\mu/\hbar\omega_{rad}$ (black points) and $k_BT/\hbar\omega_{rad}$ (red points) with radial trap frequency $\omega_{rad}/2\pi$ at $\delta f$ = 100 kHz. The horizontal dashed line in (d) represents the energy of the lowest radial vibrational excited state ($k_BT = \mu = \hbar\omega_{rad}$). Adapted from [35].

## III. DESIGN AND FABRICATION OF SUB-MICRON-PERIOD SQUARE AND TRIANGULAR MAGNETIC LATTICES

Until recently, magnetic lattices, both one-dimensional [21, 28, 33, 35] and two-dimensional [22, 24, 34], have been limited to lattice spacings ≥10 µm.

The tunnelling rate $J$ between lattice sites for a barrier height $V_0 \gg$ the tunneling rate $J$ can be expressed in terms of the lattice recoil energy $E_R = h^2/8ma^2$ by [6]

$$J/E_R \approx \frac{4}{\sqrt{\pi}}(V_0/E_R)^{3/4}\exp[-2(V_0/E_R)^{1/2}]. \tag{6}$$

For a lattice spacing $a$ = 10 µm, the tunnelling rate for Rb atoms between lattice sites is negligibly small, e.g., $J/h \approx 0.01$ Hz, or tunnelling time 16 s, for $V_0 \approx 20E_R$ (6 nK). Thus, an array of BECs trapped in a 10 µm-period lattice represents an array of isolated clouds with negligible interaction and no phase coherence between them. For a lattice spacing $a$ = 0.7 µm, the tunnelling rate for a barrier height $V_0 \approx 12E_R$ becomes $J/h \approx 17$ Hz (or tunnelling time 10 ms), which is suitable, for example, for realising the superfluid to Mott insulator quantum phase transition [11].

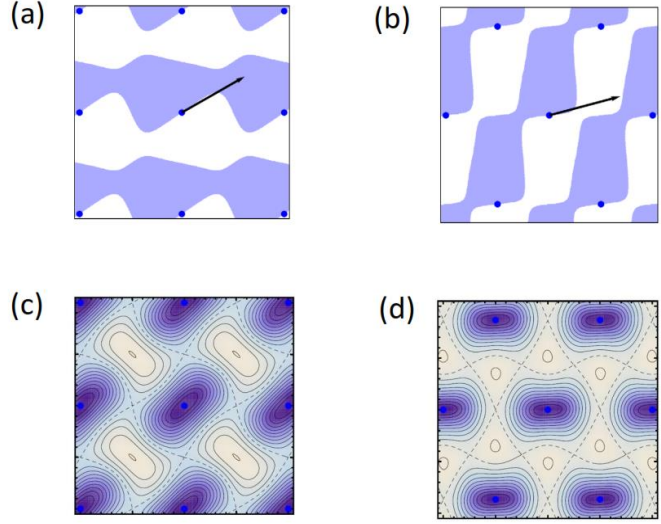

**Fig. 5.** Magnetic film patterns designed to create (a) square and (b) triangular magnetic lattices at a trapping height $z_{min} = a/2$. Blue regions represent the magnetic film; dark blue dots indicate positions of magnetic field minima; and black arrows indicate direction of the magnetic field at the minima. (c), (d) Corresponding 2D contour plots of the optimised magnetic lattice potentials for the square (c) and triangular (d) lattices with the required bias fields. Blue regions represent potential minima. Adapted from [27].

Magnetic film patterns designed to create square and triangular magnetic lattices at a trapping height $z_{min} = a/2$, using a linear programing algorithm developed by Schmied et al. [27], are shown in Figs. 5(a) and (b), along with the corresponding 2D contour plots in Figs. 5(c) and (d). The magnetic film patterns are equivalent to those produced by a virtual current circulating around the perimeter of the patterned structures, which correspond to square and triangular arrays of current-carrying Z-wire traps which have non-zero magnetic field minima.

The magnetic films used in the fabrication of 0.7 µm-period square and triangular magnetic lattice structures consist of a stack of eight bilayers of alternating Pd (0.9 nm) and Co (0.28 nm) [36, 58], with an effective magnetic film thickness $t_m$=10.3 nm [38]. These multilayer films have a large perpendicular magnetic anisotropy and a high degree of magnetic homogeneity. In addition, they exhibit square-shaped hysteresis loops [36] with large remanent perpendicular magnetisation ($4\pi M_z$ = 5.9 kG) and coercivity ($H_c \approx$ 1.0 kOe) [36], a high Curie temperature (300-400 °C) and very small grain size (~6 nm [59] compared with ~ 40 nm for TbGdFeCo [60]), allowing smooth and well-defined magnetic potentials at sub-micron lattice spacings [61]. They are also known to exhibit an enhanced magnetisation relative to bulk cobalt due to polarisation of the Pd atoms by the nearby Co layers (e.g., [62]). The Co/Pd multilayers were deposited by dc-magnetron sputtering onto a seed layer of Ta on a Si(100) substrate [36].

The magnetic microstructures were fabricated by patterning the Co/Pd multi-atomic layer magnetic film using electron-beam lithography followed by reactive ion etching. The patterned magnetic film is coated with a reflective 50 nm-layer of gold plus a 25 nm protective layer of silica. Figure 6(a) shows scanning electron microscope images of part of the 0.7 µm-period, 1 mm$^2$ triangular magnetic microstructure, which illustrates the quality of the microstructures. The patterned Co/Pd magnetic film was then glued to a direct bonded copper (DBC) [63] atom chip comprising a *U*-wire structure (for a quadrupole magnetic field) and a *Z*-wire structure (for a magnetic trap with non-zero minimum) [38]. The atom chip can accommodate four separate magnetic lattice structures (each with a patterned area of 1 mm$^2$), each of which has a *U*-wire and *Z*-wire structure directly beneath it (Fig. 6(b)). Finally, the 0.7 µm-period magnetic lattice structures are magnetised and then characterised by magnetic force and atomic force microscopy, prior to mounting in the UHV chamber.

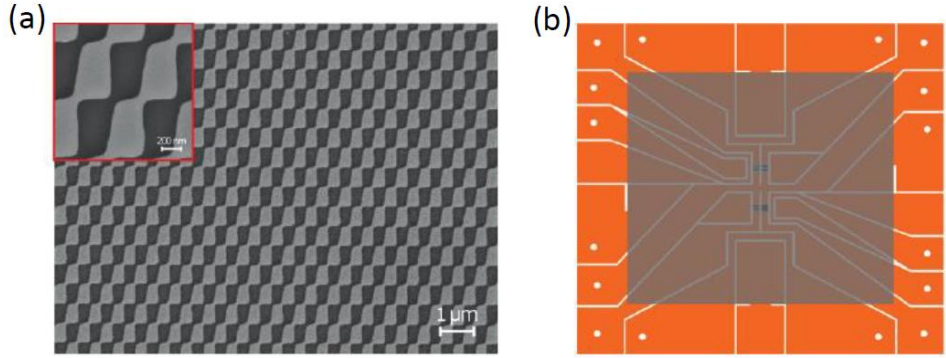

**Fig. 6.** (a) SEM image of part of the fabricated 0.7 µm-period triangular Co/Pd magnetic lattice structure. Light regions represent the (unetched) magnetic film. (b) Schematic of the direct bonded copper (DBC) atom chip, which includes four separated current-carrying *U*-wire and *Z*-wire structures for trapping and loading atoms into the magnetic lattice traps plus two wires on either side for RF evaporative cooling or RF spectroscopy. The four small green squares in the centre show the positions of the four 1 mm$^2$ magnetic lattice structures, which are located below their respective *U* and *Z*-wires. Adapted from [38].

For a triangular magnetic lattice with parameters $a = 0.7$ µm, $4\pi M_z = 5.9$ kG and $t_m = 10.3$ nm, the required bias magnetic fields for a lattice optimised for $z_{min} = a/2$ are $B_x = 0.5$ G, $B_y = 4.5$ G. The magnetic film pattern and corresponding 2D contour plot for the triangular magnetic lattice with these parameters is shown in Figs. 7(a) and (b). In the magnetic lattice trapping experiment described in Sect. IV, the triangular magnetic lattice is loaded with atoms from a *Z*-wire magnetic trap operating with bias fields $B_x = 52$ G and $B_y = 0$. Figure 7(c) shows the 2D contour plot for the 0.7-µm-period triangular lattice structure with bias fields $B_x = 52$ G, $B_y = 0$ and the above parameters. The traps for this magnetic lattice are more elongated and tighter than for the triangular lattice optimised for $z_{min} = a/2$ with $B_x = 0.5$ G, $B_y = 4.5$ G and each trap is surrounded by four rather than six potential maxima.

For a perpendicularly magnetised film structure, the magnetisation can be modelled as a virtual current circulating around the edges of the patterned structure, as indicated by the red and black arrows in Fig. 7(a). When a $B_y$ bias field is applied it can cancel the magnetic field produced by the virtual current flowing along the horizontal black edge of the patterned structure to create a periodic array of magnetic traps aligned along the short horizontal black edges [Fig. 7(b)]. On the other hand, when a $B_x$ bias field is applied it can cancel the magnetic field produced by the virtual current flowing along the vertical red edge to create a periodic array of elongated magnetic traps aligned along the long vertical red edges [Fig. 7(c)]. In general, the $B_x$ bias field for the structure in Fig. 7(c) produces lattice traps which are closer to the magnetic film, and which are tighter and deeper.

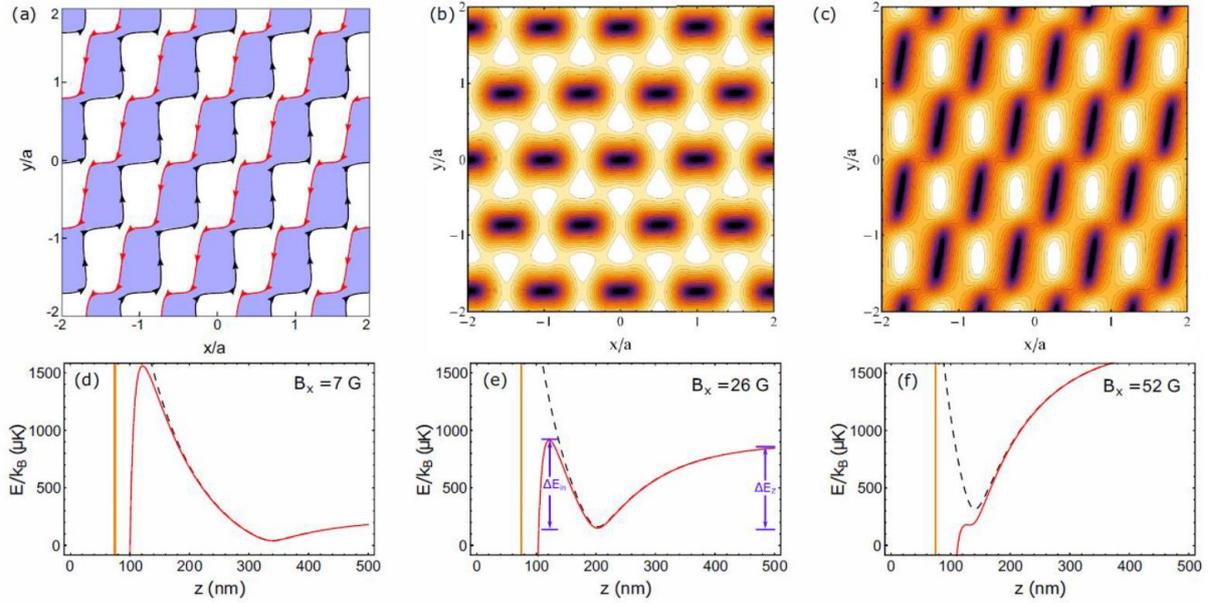

**Fig. 7.** (a) Magnetic film pattern and (b) corresponding 2D contour plot designed to create a triangular magnetic lattice optimised for $z_{min}=a/2$, with $a=0.7$ μm, $4\pi M_z=5.9$ kG, $t_m=10.3$ nm, $B_x=0.5$ G and $B_y=4.5$ G. Blue regions in (a) represent the magnetic film, dark regions in (b) are trap minima, and arrows in (a) represent virtual currents circulating around the edges of the film structure. (c) 2D contour plot of a triangular magnetic lattice potential with bias fields $B_x=52$ G, $B_y=0$, with $a=0.7$ μm, $z_{min}=139$ nm. (d)-(f) Calculated trapping potentials for $^{87}$Rb $|F = 1, m_F = -1\rangle$ atoms in a 0.7 μm-period magnetic lattice with bias fields $B_x =$ (d) 7 G ( e) 26 G, and (f) 52 G and the above parameters and surface film thickness $t_s = t_{Au} + t_{SiO2} = 75$ nm. Black dashed lines are the magnetic lattice potentials and red solid lines include the Casimir-Polder interaction with $C_4=8.2\times10^{-56}$ Jm$^4$ for a silica surface. Vertical orange lines indicate the position of the silica surface ($z=75$ nm) used in the calculations. $\Delta E_z$ and $\Delta E_{CP}$ in (e) are the barrier heights for the magnetic lattice potential only and for the magnetic lattice plus Casimir-Polder potential. Adapted from [38].

## IV. TRAPPING ULTRACOLD ATOMS IN A 0.7 μm-PERIOD TRIANGULAR MAGNETIC LATTICE

For a 0.7 μm-period magnetic lattice, the atoms are trapped at distances down to about 100 nm from the chip surface and at trapping frequencies up to about 1 MHz, which is new territory for trapping ultracold atoms. At such short distances, possible effects of surface interactions need to be considered.

The trapping potential at distance $z$ from the magnetic surface with magnetic potential $V_M(z)$ may be expressed as

$$V(z) = V_M(z) + V_{CP}(d) + V_\alpha(d), \qquad (7)$$

where $V_{CP}(d) = -C_4/[d^3(d + 3\lambda_{opt}/2\pi^2)]$ is the combined attractive Casimir-Polder (C-P) and van der Waals potential [64], $C_4$ is the C-P coefficient, $d = z_{min} - t_s$ is the distance of the trap centre to the atom chip surface allowing for a surface film thickness $t_s$, and $\lambda_{opt}$ is the wavelength of the strongest electric dipole transition of the atom. $V_\alpha(d) = -(\alpha_0/2)E_0(d)^2$ is the Stark potential arising from the interaction of the lattice trapped atoms (with polarizability $\alpha_0$) with electric fields $E_0(d)$ created by Rb atoms adsorbed onto the chip surface during each cooling and trapping sequence [46]. The attractive C-P potential can distort the repulsive magnetic potential to create a potential barrier at distances very close to the surface (Fig. 7 (d)-(f)). For bias fields $B_x < 26$ G and the parameters in Fig. 7, the calculated trap centre is located at $d > 150$ nm from the chip surface and the effect of the C-P interaction is small. For $B_x > 40$ G, the calculated trap centre is located at $d < 110$ nm and the magnetic potential is modified by the attractive C-P interaction, while for $B_x=52$ G the trapping potential becomes very shallow (trap depth ~1.5 μK). Increasing the distance to the chip surface by, for example, 25 nm increases the trap depth to 660 μK.

For the attractive Stark potential $V_\alpha(d)$ our estimates indicate that for ground-state Rb atoms [$\alpha_0(5s) = 7.9 \times 10^{-2}$ Hz/(V/cm)$^2$] interacting with typical electric fields (~900 V/cm [51]) produced by Rb atoms adsorbed on a silica surface at trap distances $d$ ~100 nm, the potential $V_\alpha(d)$ is negligibly small. However, this potential can be large for highly excited Rydberg atoms, which have a huge polarizability [e.g., $\alpha_0(30d_{3/2}) = 2.5 \times 10^6$ Hz/(V/cm)$^2$].

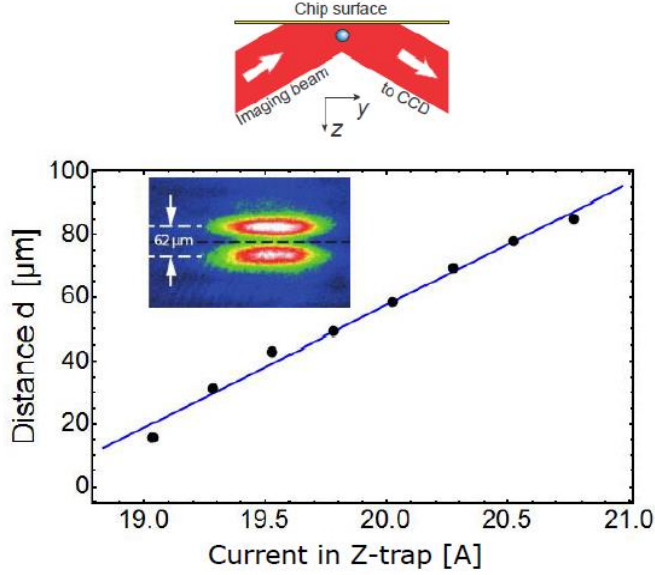

**Fig. 8.** Calibration of distance *d* of the Z-wire trap centre to the gold reflecting layer on the magnetic lattice chip surface for $B_x$=52 G versus Z-wire current $I_z$. Inset: reflective absorption image of the atom cloud close (31 μm) to the chip surface, showing the real and mirror images. Adapted from [38].

Loading of the magnetic lattice commences with ~$10^6$ $^{87}$Rb $|F = 1, m_F = -1\rangle$ atoms cooled to ~1 μK in the Z-wire trap at $d \approx 670$ μm from the chip surface with $B_x$=52 G and $I_z$=38 A. At this bias field, the loading procedure involves simply ramping down $I_z$ until the Z-wire trap potential merges with the magnetic lattice traps. Figure 8 shows a calibration of the distance *d* of the Z-wire trap centre to the gold reflecting layer on the chip surface versus Z-wire current $I_z$ as the Z-wire trapped atoms approach the chip surface. For loading with $B_x <$ 52 G, $B_x$ needs to be reduced first before loading atoms into the magnetic lattice traps and $I_z$ is reduced simultaneously to compensate for the resulting change in $z_{min}$. Next, $I_z$ is further reduced while keeping $B_x$ fixed, to allow the Z-wire trap to merge with the magnetic lattice potential at $d \approx 100$ nm from the chip surface. The ramping speed for $I_z$ is carefully optimised to prevent the Z-wire trapped atoms from penetrating the magnetic lattice potential or being lost by surface interactions or sloshing. Once the magnetic lattice is loaded, $I_z$ is increased to move the Z-wire cloud further from the surface for imaging.

In Fig. 9(a), an atom cloud is observed mid-way between two larger clouds which remains when the Z-wire trap atoms are removed either by projecting them vertically to hit the chip surface (Fig. 9(b)) or by switching off the Z-wire current. We attribute this smaller cloud to atoms trapped in the magnetic lattice, while the two larger clouds at the top and bottom are mirror and real images of atoms remaining in the Z-wire trap. The atoms trapped in individual lattice sites, which are separated by only 0.7 μm, are not resolved. Absorption measurements indicate ~2×10⁴ Rb atoms are trapped in an area of ~50×50 μm² containing about 5000 lattice sites, corresponding to $\bar{N}_{site}$≈4 atoms per site. The lifetimes of the lattice trapped atoms measured by recording the number of atoms versus hold time at different bias fields $B_x$ range from 0.4 ms to 1.7 ms and increase slowly with distance $d$ from the chip surface (Fig. 10(a)).

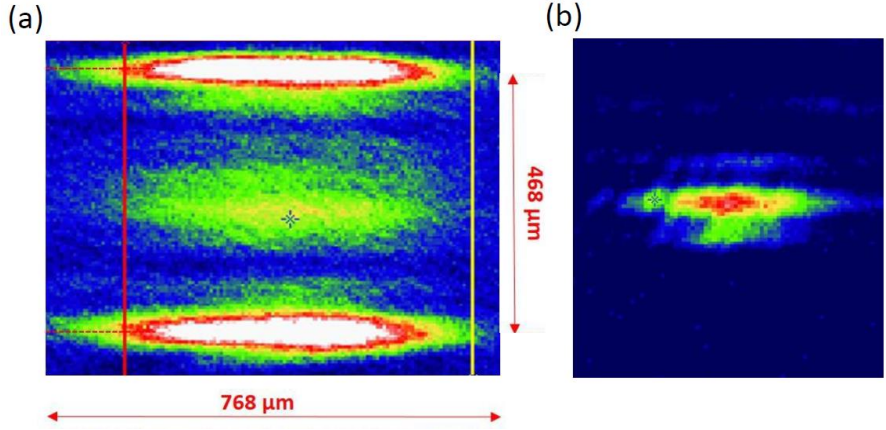

**Fig. 9.** Reflection absorption images of ⁸⁷Rb $|F = 1, m_F = -1\rangle$ atoms (a) trapped in the 0.7 μm-period triangular magnetic lattice mid-way between the real and mirror images of the Z-wire trapped cloud, for $B_x$ = 52 G; (b) trapped in the 0.7 μm-period triangular magnetic lattice only, for $B_x$ = 13 G Adapted from [65].

To interpret the short trapped atom lifetimes and their dependence on distance from the surface, we consider several possible loss processes. When the atoms are transferred from the Z-wire trap ($\bar{\omega}/2\pi$ ≈100 Hz) to the very tight magnetic lattice traps ($\bar{\omega}/2\pi$ ≈ 300-800 kHz) the resulting compression is estimated to heat the cloud from ~1 μK in the Z-wire trap to 3-8 mK in the magnetic lattice traps. During this compression, atoms with energies higher than the effective trap depth in the *z*-direction rapidly escape the lattice traps, resulting in a sudden truncation of the high energy tail of the Boltzmann energy distribution. The remaining more energetic atoms in the outer region of the magnetic lattice traps with energies comparable to the trap depth can overcome the trap barrier and are rapidly lost or spill over into neighbouring lattice sites. The remaining trapped atoms reach a quasi-equilibrium at a lower temperature determined by the truncation parameter η ≈ $\Delta E_{eff}/k_BT$, where $\Delta E_{eff}$=Min{$\Delta E_z$, $\Delta E_{CP}$} (Fig. 8(e)). The evaporation loss rate is rapid at the beginning of the evaporation and then decreases as evaporation progresses.

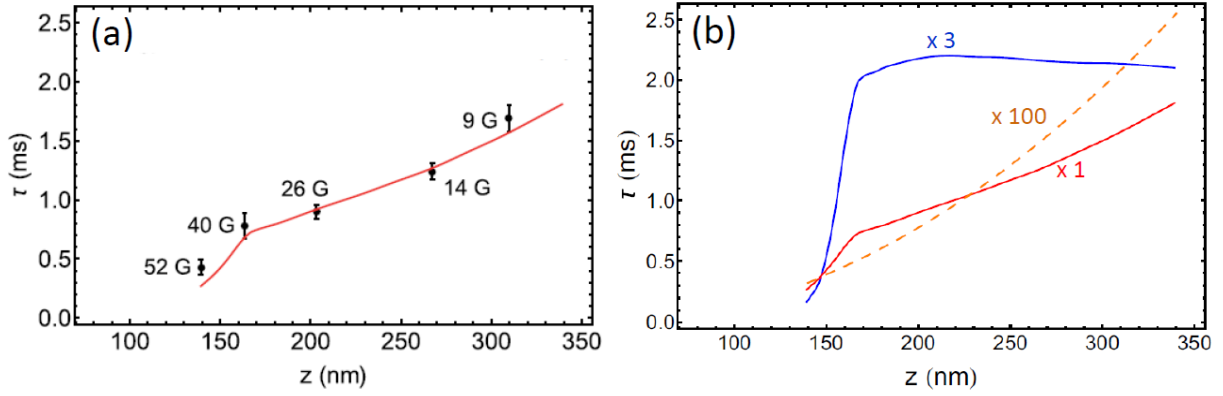

**Fig. 10.** (a) Measured lifetimes (black points) of atoms trapped in the 0.7 μm-period triangular magnetic lattice versus distance $z$ of the lattice trap centre from the magnetic film surface. The red curve shows calculated evaporation lifetimes $\tau_{ev}$ for $\eta = 4$, $\bar{N}_{site} = 1.5$, offset $\delta z = 25$ nm and the fixed parameters given in Fig. 7 caption. (b) Calculated total lifetimes for evaporation $\tau_{ev}$ [red (second) curve], three-body recombination $\tau_{3b}$ [blue (top) curve] and spin flips $\tau_s$ (dashed orange curve). The chip surface is located at $z=50$ nm. The curves for $\tau_{3b}$ and $\tau_s$ are reduced by factors of 3 and 100, respectively. Adapted from [38].

Using a 1D evaporation model [45], the lifetime for 1D thermal evaporation can be expressed as

$$\tau_{ev} = \tau_{el}/[f(\eta)e^{-\eta}], \qquad (8)$$

where $\tau_{el} = [n_0 \sigma_{el} \bar{v}_{rel}]^{-1}$, $n_0 = \bar{N}_{site}/[m/(2\pi k_B T)]^{3/2} \bar{\omega}^3$ and $f(\eta) \approx 2^{-5/2}[1-\eta^{-1}+1.5\eta^{-2}]$. According to this model, $\tau_{ev}$ scales as $\Delta E_{eff}/[\bar{\omega}^3 \bar{N}_{site} \eta f(\eta) e^{-\eta}]$. For decreasing bias fields $B_x < 40$ G (where $\Delta E_{eff} \equiv \Delta E_z$), the trap minima move away from the chip surface and $\bar{\omega}^{-3}$ increases faster than $\Delta E_{eff}$ decreases, so that $\tau_{ev}$ exhibits a slow almost linear increase with increasing distance $z$ (Fig. 10(b)). For increasing $B_x \geq 40$ G (where $\Delta E_{eff} \equiv \Delta E_{CP}$), the trap minima move very close to the chip surface and $\Delta E_{eff}$ and $\bar{\omega}^{-3}$ both decrease together, resulting in a sharp decrease in $\tau_{ev}$.

A second possible loss process is three-body recombination in the tight magnetic lattice traps, for which the lifetime is given by $\tau_{3b} = 1/(K_3 n_0^2)$, where $K_3 = 4.3(1.8) \times 10^{-29}$ cm$^6$s$^{-1}$ for non-condensed $^{87}$Rb $|F = 1, m_F = -1\rangle$ atoms [50, 51], so that $\tau_{3b}$ scales as $\Delta E_{eff}^3/[\bar{\omega}^6 \bar{N}_{site}^2 \eta^3]$. For decreasing $B_x < 40$ G (where $\Delta E_{eff} \equiv \Delta E_z$), $\Delta E_{eff}^3$ decreases at about the same rate as $\bar{\omega}^{-6}$ increases, resulting in $\tau_{3b}$ remaining almost constant for distances $z > 200$ nm (Fig. 10(b)). For increasing $B_x \geq 40$ G (where $\Delta E_{eff} \equiv \Delta E_{CP}$), the trap minima move very close to the chip surface and $\Delta E_{eff}^3$ and $\bar{\omega}^{-6}$ both decrease strongly with decreasing $z$, resulting in a rapid decrease in $\tau_{3b}$ (Fig. 10(b)). A further possible loss process can result from spin flips due to magnetic Johnson noise generated by the gold conducting layer near the surface of the magnetic film [45, 47, 48]. For a gold conducting layer of thickness $t_{Au}$, $\tau_s \approx 0.13 \left( d' + \frac{d'^2}{t_{Au}} \right)$ ms [38], where $d' = z_{min} - t_{Au}$ and $t_{Au}$ are in nanometres. The calculated spin-flip lifetimes, which range from $\tau_s = 46$ ms for $d' = 110$ nm to 240 ms for $d' = 310$ nm, are much longer than the measured lifetimes.

The calculated evaporation lifetime $\tau_{ev}$ versus distance (red curve, Fig. 10(b)) has a positive slope given by $\Delta E_{eff}/(\overline{\omega}^3 d)$ that closely matches the slope of the measured lifetime versus distance, whereas the calculated three-body loss lifetime $\tau_{3b}$ versus distance remains almost constant for distances $z > 200$ nm. The red curve in Fig. 10(a) shows the calculated evaporation lifetime $\tau_{ev}$ with fitted scaling parameters $\eta=4$ and $\overline{N}_{site}=1.5$, a fitted offset $\delta d = 25$ nm (see below) and the fixed parameters given in the Fig. 7 caption. The smaller value $\overline{N}_{site}=1.5$ compared with the $\overline{N}_{site}\approx 4$ estimated from the total number of atoms trapped in ~5000 lattice sites could be a result of atoms spilling over into neighbouring lattice sites during the initial transfer of atoms into the tight magnetic lattice traps. A value of $\overline{N}_{site}\approx 1.5$ is characteristic of an end product of three-body recombination during the earlier equilibrating stage when the atom densities are very high. To obtain a reasonable fit at distances $d < 100$ nm from the chip surface, where the lifetime is very sensitive to the $d^{-4}$ dependence of the C-P interaction, requires either the calculated $C_4=8.2\times 10^{-56}$ Jm$^4$ to be an order of magnitude smaller, which is unrealistic, or the calculated distances to be slightly larger, by $\delta d \approx 25$ nm, which is within the estimated uncertainty in $d = z_{min} - (t_{Au} + t_{SiO_2})$ [38].

The above results suggest that the atom lifetimes in the 0.7 µm-period magnetic lattice are currently limited mainly by losses due to evaporation following transfer of the atoms from the Z-wire trap into the very tight magnetic lattice traps, rather than by losses due to fundamental processes such as surface interactions, three-body recombination or spin flips caused by magnetic Johnson noise.

The measured lifetimes of the atoms trapped in the 0.7 µm-period magnetic lattice, 0.4 - 1.7 ms, need to be increased significantly to allow RF evaporative cooling in the magnetic lattice and to allow quantum tunnelling, where the relevant tunnelling times for a 0.7 µm-period lattice are typically ~10 ms at a barrier height $V_0 \approx 12 E_R$. The biggest increase in trap lifetime is likely to come from improving the transfer of atoms from the Z-wire trap to the very tight magnetic lattice traps. Heating due to adiabatic compression during transfer of the atoms could be reduced by loading from a magnetic trap with large radial trap frequency, such as the type reported by Lin et al. ($\omega_{rad}/2\pi \approx 5$ kHz) [45] or from a 1D optical lattice of pancake traps ($\omega_{rad}/2\pi \approx 100$ kHz) [66].

## V. LONG-RANGE INTERACTING RYDBERG ATOMS IN A LARGE-SPACING MAGNETIC LATTICE

Highly excited Rydberg atoms can be orders of magnitude larger than ground-state atoms, making them extremely sensitive to one another and to external fields [49]. At large atom separations, Rydberg-Rydberg interactions involve long-range van der Waals (vdW) interactions which scale as $n^{11}$ with the principal quantum number $n$ [49]. Thus atoms on neighbouring sites of a large-spacing magnetic lattice can interact by exciting the atoms to Rydberg states [29, 34, 50, 51].

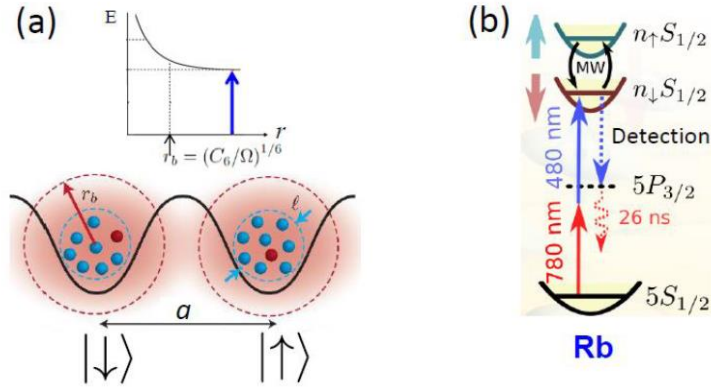

**Fig. 11.** Scheme for creating long-range spin-spin interactions between single Rydberg atoms trapped in neighbouring sites of a large-spacing magnetic lattice. (a) Two spin states $|\downarrow\rangle$ and $|\uparrow\rangle$ are encoded in a single two-photon excitation to the Rydberg state $|n_\downarrow S\rangle$ or $|n_\uparrow S\rangle$ (red spheres), which is shared amongst all atoms in the ensemble (blue spheres). To prepare a single Rydberg atom in an ensemble of spatial extent $l$ on each site of a lattice with period $a$ requires $l \ll r_b \lesssim a$, where $r_b \approx |C_6/\Omega|^{1/6}$ is the Rydberg blockade radius. (b) Level structure of a single atom involving two-photon excitation to the Rydberg $|n_\downarrow S\rangle$ state. The states involved in the detection processes are marked with dotted lines. Adapted from [50].

We consider a magnetic lattice in which each lattice site $i$ contains an ensemble of $N_i$ rubidium-87 atoms of spatial extension $l$ and the different sites are separated by the lattice period $a$ (Fig. 11(a)). Each lattice site is prepared with precisely one Rydberg excitation, for example, by tuning to the two-photon laser excitation $|g\rangle \rightarrow |R^\downarrow\rangle \equiv |n_\downarrow S_{1/2}, m_j=+1/2\rangle$ (Fig. 11(b)). To restrict the system to a single excitation on each lattice site we make use of Rydberg blockade in which the presence of the Rydberg atom shifts the energy levels of nearby atoms, thereby suppressing subsequent excitation of other atoms in the ensemble (Fig. 11(a)) [49]. The characteristic range of the Rydberg-Rydberg interaction is given by the Rydberg blockade radius, $r_b \approx |C_6/\Omega|^{1/6}$, which for a typical atom-light coupling constant $\Omega/2\pi \approx 1$ MHz is 2 - 10 μm, depending on the Rydberg state [50]. In order to prepare a single Rydberg atom in each lattice site separated by a distance $a$ we require a separation of length scales $l \ll r_b \lesssim a$, which can be met for a large-spacing magnetic lattice. The use of atomic ensembles avoids the problem of exact single-atom filling of magnetic lattice sites and single-atom detection, and also helps with the initialisation and readout of individual Rydberg spin states.

To initialise the spin lattice we use collectively enhanced atom-light coupling in each microtrap to drive two-photon Rabi oscillations between the ground state and the Rydberg state. Complete population inversion can be achieved by exciting with a Rabi π-pulse. Numerical simulations [50] indicate that, for a typical magnetic lattice microtrap with radii $\sigma_x = \sigma_y = 0.15$ μm, $\sigma_z = 0.4$ μm, the optimal principal quantum number for maximising the efficiency of initial state preparation for $N = 5$-15 atoms per site is around $n \approx 33$, with an estimated overall efficiency of around 92%, which is limited mainly by atom number fluctuations due to the stochastic loading process.

Following initialisation, the excitation laser is switched off and Rydberg excitations on neighbouring lattice sites interact as a result of their giant electric dipole moments (typically several kilodebye). We identify two collective spin states [50]

$$|\uparrow\rangle = \frac{1}{\sqrt{N}}\sum_j |g_1,\ldots,g_{j-1},R^\uparrow,g_{j+1},\ldots,g_N\rangle \tag{9a}$$

$$|\downarrow\rangle = \frac{1}{\sqrt{N}}\sum_j |g_1,\ldots,g_{j-1},R^\downarrow,g_{j+1},\ldots,g_N\rangle, \tag{9b}$$

where $|R^\uparrow\rangle$ and $|R^\downarrow\rangle$ denote the $|n_\uparrow S_{1/2}, m_j=+1/2\rangle$ and $|n_\downarrow S_{1/2}, m_j=+1/2\rangle$ Rydberg states. These collective spin states are coherent superpositions with the single Rydberg excitation shared amongst all atoms in the ensemble [49]. This configuration allows complex spin-spin interactions including XXZ spin-spin interactions in 2D. In addition, the two collective spin states may be coupled using two-photon microwave transitions between the two Rydberg states (Fig. 11(b)) to realise single-spin rotations which simulate transverse and longitudinal magnetic fields.

For the quantum simulation of spin models, the spin-spin coupling rate between neighbouring lattice sites $|C_6|/a^6$ needs to greatly exceed the decoherence rate Γ of the Rydberg state. In addition, in order to prevent interference between neighbouring sites during the initialisation phase, the Rydberg excitation bandwidth $\sqrt{N}\Omega$ needs to exceed the spin-spin coupling rate between neighbouring sites. Finally, to ensure good conditions for Rydberg blockade, the spin-spin coupling rate between atoms within each ensemble $|C_6|/l^6$ needs to greatly exceed the excitation bandwidth. These constraints lead to [50]:

$$\frac{|C_6|}{l^6} \gg \sqrt{N}\Omega \gg \frac{|C_6|}{a^6} \gg \Gamma. \tag{10}$$

For example, for Rydberg $33S_{1/2}$ atoms ($\Gamma/2\pi$=7.3 kHz [67]) trapped in a magnetic lattice with period $a\approx$2.5 μm, trap size $l\approx 2\sigma$=0.8 μm, $N$=10 atoms per site and typical collectively-enhanced 2-photon Rabi frequency ($\sqrt{N}\Omega/2\pi\approx$3 MHz), each of the criteria in (9) is satisfied by at least an order of magnitude.

To read out the collective spin state we need to detect the presence of a single Rydberg atom in a given spin state in the atomic ensemble with high fidelity. This can be achieved by using a single-Rydberg atom triggered ionisation 'avalanche' scheme [68, 69], in which the presence of the single Rydberg atom conditionally transfers a large number of ground-state atoms in the trap to untrapped states which can then be detected by standard site-resolved absorption imaging [50].

We now consider the realisation of lattice spin models where the spin-1/2 degree of freedom is encoded in the collective spin states of Eq. (8). By using two Rydberg $S$-states with different principal quantum numbers, i.e., $|R^{\uparrow}\rangle = |n_{\uparrow}S_{1/2}, m_j=+1/2\rangle$ and $|R^{\downarrow}\rangle = |n_{\downarrow}S_{1/2}, m_j=+1/2\rangle$ with $n_{\uparrow} \neq n_{\downarrow}$, we can realise a spin-1/2 exchange Hamiltonian, with spin-spin couplings of the form [70]

$$H = \sum_{i,j<i}[J_z(r_{ij})S_i^z S_j^z + \frac{1}{2}J_\perp(r_{ij})(S_i^+ S_j^- + (S_i^- S_j^+)] + \sum_i[z_i \tilde{h}_\parallel S_i^z + h_\perp S_i^x - h_\parallel S_i^z]. \tag{11}$$

Here, $S_i^z$ denotes the $z$-component of the spin-1/2 operator and $S_i^\pm$ is the spin raising/lowering operator at lattice site $i$, $J_z$ and $J_\perp$ are spin-coupling coefficients that originate from the vdW interactions between the chosen spin states, and $h_\parallel$ and $h_\perp$ are longitudinal and transverse fields. The inclusion of a tunable microwave field to couple the $n_\downarrow S_{1/2} \leftrightarrow n_\uparrow S_{1/2}$ states via a two-photon microwave transition results in tunable longitudinal and transverse field terms $h_\perp$ and $h_\parallel$. For $nS$ states, the spin-coupling coefficients are effectively isotropic and depend only on the distance between the two sites $i$ and $j$, i.e., $J_z(r_{ij}) = J_z/|r_i - r_j|^6$ and $J_\perp(r_{ij}) = J_\perp/|r_i - r_j|^6$. For the Ising spin-coupling coefficient we obtain $J_z = C_6(n_\uparrow, n_\uparrow) + C_6(n_\downarrow, n_\downarrow) - 2C_6(n_\uparrow, n_\downarrow)$, where the $C_6(n_1, n_2)$ coefficients denote the diagonal vdW interaction between the Rydberg states $|n_1 S_{1/2}, 1/2\rangle \otimes |n_2 S_{1/2}, 1/2\rangle$. The $J_\perp = 2\tilde{C}_6(n_\uparrow, n_\downarrow)$ term arises as an exchange process between the degenerate states $|n_\downarrow S_{1/2}, 1/2\rangle \otimes |n_\uparrow S_{1/2}, 1/2\rangle$ and $|n_\uparrow S_{1/2}, 1/2\rangle \otimes |n_\downarrow S_{1/2}, 1/2\rangle$ via

vdW interactions and depends strongly on $\delta n = n_\uparrow - n_\downarrow$. The longitudinal field $\tilde{h}_\parallel = [C_6(n_\uparrow, n_\uparrow) - C_6(n_\downarrow, n_\downarrow)]/2$ originates from the small difference between intra-spin interactions where $z_i = \sum_j r_{ij}^{-6}$ is a factor depending on the lattice geometry.

The Hamiltonian (11) allows studies of anisotropic XXZ spin-1/2 models in various geometries with additional longitudinal and transverse fields. These models can allow the study of exotic quantum phases of matter [71]. A key parameter is the anisotropic ratio $\Delta = J_z/J_\perp$ which can be tuned over a wide range by choosing the principal quantum numbers $n_\uparrow$ and $n_\downarrow$. The case of no transverse field $h_\perp$ has been studied extensively in *one-dimensional* spin chains with next-neighbour interactions and supports ferromagnetic ($\Delta < -1$) and antiferromagnetic ($\Delta > 1$) phases as well as a spin liquid phase for $-1 < \Delta < 1$. For the anisotropic XXZ spin-1/2 model in *two dimensions*, no exact solution exists; however it is expected to support non-trivial quantum phases which depend on the lattice geometry. For example, for a triangular lattice it gives rise to a stable supersolid phase [72, 73] while for a kagome lattice a spin-singlet valence-bond solid phase emerges [74-76]. In addition, due to the $1/r^6$ character of the spin-spin interactions one can realise frustrated $J - J'$ models on a square (or rhombus) lattice which are expected to yield stable stripe-like supersolid phases [70, 77].

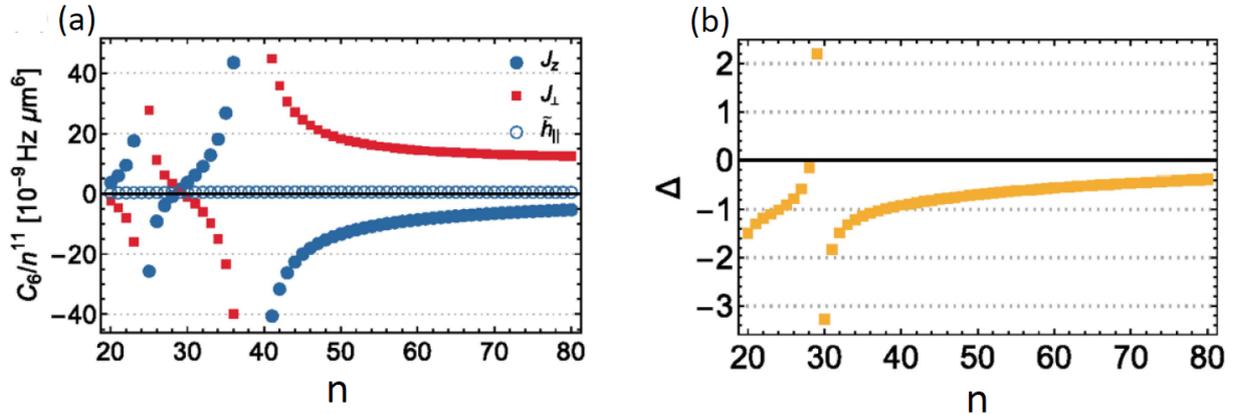

**Fig. 12.** (a) Calculated spin coefficients of the Hamiltonian (11) as a function of principal quantum number $n$ for the Rb $nS_{1/2}$ and $(n+1)S_{1/2}$ states. (b) Resulting anisotropy ratio $\Delta = J_z/J_\perp$. Förster resonances occur at around $n \approx 24$ and $n \approx 38$. Adapted from [50].

To illustrate the tunability of the resulting spin interactions we consider the case $n_\uparrow = n_\downarrow + 1$ for which the two Rydberg states are close in energy and the exchange process $J_\perp$ is maximised [50]. Figure 12 shows calculations of (a) the coupling strengths for Rb $nS_{1/2}$ and $(n+1)S_{1/2}$ states, and (b) the resulting anisotropy parameter $\Delta = J_z/J_\perp$, as a function of the principal quantum number $n$. Both $J_z$ and $J_\perp$ exhibit two Förster resonances, at $n \approx 24$ and $n \approx 38$, where the channels to $\{nP_{3/2}, nP_{1/2}\}$ and $\{nP_{3/2}, nP_{3/2}\}$ states become close in energy, respectively. As a result of these Förster resonances it is possible to realise ferromagnetic $J_z$ interactions for $n \in \{25, 28\}$ and $n > 38$ or antiferromagnetic spin interactions for $n < 25$ and $n \in \{29, 38\}$. The anisotropy parameter $\Delta$ crosses the transition from ferromagnetic to spin-liquid phases at $n = 40$ ($\Delta = -1$). The inclusion of a tunable microwave field allows additional control, including time-dependent control of the transverse and longitudinal fields. In Appendix C of Ref. [48] we also considered spin encoding using two different Rydberg $nP$ states, which can give rise to even richer spin Hamiltonians with anisotropic couplings such as generalised compass type models.

The time-scales associated with the atomic motion (~ ms) or lifetimes of high $nS$ Rydberg states (> 20 μs) [67] are long compared with the time-scales associated with strong Rydberg-Rydberg interactions (~1 μs). This enables investigation of non-equilibrium spin dynamics on both short and long times, including, for example, the build-up of spin-spin correlations following a sudden quench of the system parameters.

A potential issue with the use of long-range interacting Rydberg atoms in magnetic lattices is the effect of stray electric fields created by Rb atoms adsorbed on the chip surface during each cooling and trapping sequence [51, 52, 78, 79]. Studies of Rb Rydberg atoms trapped at distances down to 20 μm from a gold-coated chip surface have revealed small distance-dependent energy shifts of ~ ±10 MHz for $n \approx 30$ [51]. Significantly larger electric fields have since been found when the chip surface is coated with a dielectric $SiO_2$ layer [52]. Recent studies have demonstrated that the stray electric fields can be effectively screened out by depositing a uniform film of Rb over the entire chip surface [78] or by using a smooth monocrystalline quartz surface film with a monolayer of Rb absorbates [79].

## VI. SUMMARY AND OUTLOOK

In this review we have discussed recent advances in our laboratory on the development of magnetic lattices created by patterned magnetic films to trap periodic arrays of ultracold atoms.

Multiple Bose-Einstein condensates of $^{87}$Rb atoms have been produced in a 10 μm-period 1D magnetic lattice, with low atom temperatures (~ 0.16 μK), high condensate fractions (~ 81%) and a high degree of lattice uniformity. For large radial trap frequencies (> 10 kHz), the clouds of trapped atoms enter the quasi-1D regime. For this 10 µm-period magnetic lattice, the elongated clouds of ultracold atoms represent arrays of isolated atomic clouds with no interaction between neighbouring sites. To achieve interaction between sites via quantum tunnelling, magnetic lattices with sub-micron periods are required.

High-quality triangular and square magnetic lattice structures with a period of 0.7 µm have been fabricated by patterning a Co/Pd multi-atomic layer magnetic film using electron-beam lithography and reactive ion etching. Ultracold atoms have been successfully trapped in the 0.7 µm-period triangular magnetic lattice at distances down to about 100 nm from the chip surface. The lifetimes of the lattice trapped clouds (0.4 - 1.7 ms) increase with distance from the chip surface. Model calculations suggest that the trap lifetimes are mainly limited by losses due to thermal evaporation following transfer of atoms from the Z-wire magnetic trap to the very tight magnetic lattice traps, rather than by fundamental loss processes such as surface interactions, three-body recombination or spin flips due to magnetic Johnson noise. It should be possible in future to improve the transfer of atoms from the Z-wire trap to the very tight magnetic lattice traps, for example, by loading the atoms from a magnetic trap or 1D optical lattice with high trap frequencies.

We have proposed an alternative approach to create interactions between atoms on neighbouring sites of a magnetic lattice which is based on long-range interacting Rydberg atoms. Each spin is encoded directly in a collective spin state involving a single $nS$ or $(n+1)S$ Rydberg atom prepared via Rydberg blockade in an ensemble of ground-state rubidium atoms. The Rydberg spin states on neighbouring lattice sites are allowed to interact via van der Waals interactions with the driving fields turned off. They are then read out using a single-Rydberg atom triggered photoionisation avalanche scheme. The use of Rydberg states provides a way to realise complex spin models including XXZ 2D spin-1/2 models. This paves the way towards engineering exotic spin models, such as those based on triangular-based lattices which can give rise to a rich quantum phase structure including frustrated-spin states. With the use of Rydberg atoms, it should also be possible to investigate dynamics such as the build-up of spin-spin correlations on different length and time scales following a dynamical change in the system [80].

**ACKNOWLEDGEMENTS**

We thank various members of the magnetic lattice group who have contributed to this project over the years, including Mandip Singh, Saeed Ghanbari, Smitha Jose, Leszek

Krzemien, Prince Surendran, Ivan Herrera and Russell McLean. We also thank Manfred Albrecht and Dennis Nissen from the University of Augsburg for providing the Co/Pd magnetic films and Amandas Balcytis, Pierette Michaux and Saulius Juodkazis for fabricating the magnetic microstructures. We thank the American Physical Society for permission to reproduce Figs. 2-4, 6-8, 10 and the Institute of Physics Publishing for permission to reproduce Figs. 5, 11, 12. We gratefully acknowledge funding from the Australian Research Council (Discovery Project DP130101160).